%% file: Bodynets_short.tex
\documentclass{sig-alternate}

\usepackage{amsfonts}
\graphicspath{{./fig/}}
\usepackage{url}
\usepackage[utf8]{inputenc}

\usepackage{tabularx}

\newlength\figurewidth

\usepackage{tikz}
\usepackage{pgf}
\usepackage{pgfplots}

\usepackage{epstopdf}

\usetikzlibrary{decorations}
\usetikzlibrary{arrows}
\usetikzlibrary{shapes.misc}
\usetikzlibrary{positioning}
\usetikzlibrary{calc}
\tikzstyle{itnode}=[circle,thick,draw]
\tikzstyle{itnodesq}=[rectangle,thick,text centered,text width=5em,minimum height=3em,draw]
\tikzstyle{init} = [pin edge={<-,thin,black}]

\usepackage{caption}
\usepackage{subcaption}
\makeatletter
\let\@copyrightspace\relax
\makeatother

\begin{document}

\CopyrightYear{2014}

\conferenceinfo{Bodynets}{2014 Edmonton, Canada}

\title{Live Group Detection for Mobile Wireless Sensor Networks}

\numberofauthors{4}
\author{
\alignauthor
Matthieu Lauzier,\\
Tanguy Risset\\
       \affaddr{Universit\'e de Lyon, INRIA}\\
       \affaddr{INSA-Lyon, CITI-INRIA, }\\
       \affaddr{F-69621, Villeurbanne, France}\\
       \email{matthieu.lauzier@insa-lyon.fr}
\alignauthor
Antoine Fraboulet\\
       \affaddr{HIKOB SA}\\
       \affaddr{F-69603, Villeurbanne, France}\\
       \email{antoine.fraboulet@insa-lyon.fr}
\alignauthor
Jean-Marie Gorce \\
       \affaddr{Princeton University, NJ, USA}\\
       \affaddr{Universit\'e de Lyon, INRIA}\\
       \affaddr{INSA-Lyon, CITI-INRIA, }\\
       \affaddr{F-69621, Villeurbanne, France}\\
       \email{jean-marie.gorce@insa-lyon.fr}
}

\maketitle

\begin{figure}[b!]
\begin{center}
\setlength{\unitlength}{1cm}
\begin{picture}(10,7) 
\end{picture}
\end{center}
\end{figure}

\begin{abstract}
This paper deals with distributed algorithms for  monitoring the topology of a dynamic group of mobile wireless sensor networks.
We propose two major extensions of a distributed static group consensus algorithm  and an experimental implementation.
Group consensus algorithms are exploited to let each node obtain the knowledge of its connected components.
The proposed extensions provide a more accurate information about the proximity of nodes and allow to deal with dynamic networks using a periodical reevaluation of the group detection. We validate these algorithms by implementing them in  an original and challenging application scenario, in the context of a real bicycle race. The real traces thus obtained and analyzed show the effectiveness of our live group detection implementation.
\end{abstract}



\section{Introduction}
Distributed decisions within any group of agents, is a very active research area and theoretical results as well as efficient algorithms have already been proposed \cite{Boyd,Gossip,Gossip2}.
In the context of wireless networks, the task is made harder due to possible transmission errors, channel asymmetry, dynamic behaviour of the channel and node mobility~\cite{kar2009distributed,marechal2010joint}.

In this paper, we consider a group of mobile agents moving roughly in a common direction. We propose algorithmic solutions allowing each agent to periodically discover its neighbours: one-hop neighbours  as well as multi-hop neighbours.
The reference scenario is a bike race, during which groups are susceptible to split or merge. The objective is a live gathering of information about who is present in a group for live TV broadcasting. For that, we need a fully distributed approach allowing every agent to discover with a consensus algorithm  the list of neighbours participating to the same pack.
This study may be of interest for various other applications such as group navigation support in crowded environments, autonomous navigation of a fleet of robots\ldots

This problem exhibits some similarities with a clustering problem. However, a clustering problem aims at exploiting the structure of a graph and to form some subgroups to ensure a good structure of the network for further communications while our objective is rather to estimate the groups naturally formed in the real world.
Hence, we have  focused on distributed decision algorithms, which are widely present in the literature. In this context, gossip approaches are very appealing \cite{Cortes2008,Boyd,Gossip,Gossip2}.
While some works focused on scaling issues to ensure a proper behaviour when the size of the network grows \cite{kermarrec2003probabilistic}, other works focused on the consensus accuracy and convergence speed \cite{Boyd,Gossip}.
Max-consensus problem has been much less studied than average consensus. In \cite{Cortes2008}, the max-consensus is mentioned as one of the possible consensus operations, but the paper doesn't provide specific results about the convergence rate. The most relevant previous contribution is provided by Iutzeler et al. in~\cite{Iutzeler2012}.

The proposed algorithms are based on the N-dimension generalization of the Random Broadcast Max-Consensus algorithm given in~\cite{Iutzeler2012}, allowing each agent to build and share the list of its muli-hop neighbors.
We extend this approach to a dynamic context where the group information needs to be updated according to possible group merge or split. 


Experimental validation has been done in the context of a cycling race with 10 agents, equipping each bicycle with a wireless sensor node to assess the interactions between the racers and to provide a live monitoring of the dynamic evolution of the cyclists groups that form during the race.
We were also able to store the data on the nodes to obtain an accurate database on the network behavior.


The rest of this paper is organised as follows. We present our mobile group consensus algorithm in section~\ref{sec-algo}. This experiment and the results are presented in section~\ref{sec-Exp} where  we provide practical evidence of the efficiency of our Algorithm.

\section{Consensus Algorithms for Mobile Groups}
\label{sec-algo}
\input{staticConsensus}

\subsection{Group Mate Consensus}
\label{sec-extension}
\label{sec:group_det}

\newcommand{\gmc}{\sc gmc}


Let's now consider that a node may not take into account the information it receives from a neighbour if a condition between the transmitter and the receiver is not satisfied (e.g. neighbour further than a threshold distance, sporadic connection...) as if the two nodes were not connected. We thus define two nodes as being \emph{mates} if the application condition is satisfied or if the two nodes have common mates.



The artificial separation of the graph using a proximity criterion (RSSI-based in practice) between the nodes is included in the full algorithm (Algorithm \ref{algo3}).
The $B^v$ vectors have now integer values (rather than boolean):  $B^v_i \in[0;M], M\in \mathbb{N}$, which represent how {\em close} nodes are with respect to the chosen criterion. 
When the $v$ node receives a message from node $j$, it updates its $B^v$ vector only if $v$ and $j$ are mates.
The set of edges $E_C$ associated to the new adjacency matrix $C$  models the partition of the graph. We also define $\Delta_C$ the diameter of the graph given by the adjacency matrix $C$.

\addtocounter{equation}{-1}

%


\newcommand{\mgmc}{\sc mgmc}
\subsection{Extension for Node Mobility }
\label{sec:period}
When the nodes move, a connected component may break down in two components, or inversely, two independent components may merge. Both are referred to as merge-and-split variations. The mobile network can be modelled as a dynamic graph $G(t)=(V,E_C(t))$, but the timescale of these variations is assumed greater than the algorithm convergence time.
The merging feature is natural with {\sgc} since a new node entering a group is eligible to send its vector which naturally propagates over the group.
But, on the contrary, if a node disappears from $C_i$, the \emph{Max} operation cannot propagate this withdrawal.
Dynamically forgetting a node would require to share more complex information, thus we rather introduce a periodical reset of all vectors $B^v$.

The main issue with this approach is the need of a synchronisation method to implement these periodical resets. Let us introduce a global network clock $K$, called {\em epoch} and for each node $v$ a local epoch indicator $k_v \in \mathbb{N}$, these clocks are virtual clocks representing an epoch stamps and not to be mistaken with hardware clocks. All local epochs $k_v$ are initialised to 0 and indicate the current {\em epoch}. These epochs are transmitted with $B^v$ vectors. 
The synchronisation consists in broacasting a new epoch indicator: in our case an external source periodically broadcasts an incremented epoch $K+1$, which is multi-hop propagated over the network using a similar max-consensus method among the nodes.
The Mobile Group Mate Consensus Algorithm ({\mgmc}) is described in Algorithm \ref{algo3}.
\begin{algorithm}
\label{algo3}
{\sc Mobile Group Mate Consensus}
\begin{tabbing}
==\= ==\= ==\= ==\= ==\= ==\kill
(Executed on each node $v$)\+ \\
Initialize $B^v$ with $M$ at component $v$ and 0 elsewhere\\
Initialize $k_v$ to 0\\
{\bf repeat} infinitely\+\\
$T_{max}=random(0-T)$\\
{\bf While}  not expired backoff $T_{max}$ \+ \\
receive packet $P$\\
{\bf If} $P=\{k_j,B^j\}$ from node $j$ \+\\
proxim=$f$({\sc rssi})  // proxim between 0 and M\\
{\bf If} $k_j>k_v$\+ \ \ \ //change epoch\\
$k_v = k_j$\\
set $B^v$ to its initial value \-\\
{\bf end If}\\
{\bf If} (proxim$>$Threshold)  \ \ //$j$ and $v$ are mates\+ \\
$B^v$=$Max(B^v,B^j))$\-\\
{\bf else} \+\\
$B^v_j = Max(B^v_j, \mbox{proxim})$\- \\

{\bf end If}\-\\
{\bf If} $P$ is beacon with epoch $K$\+\\
{\bf If} $K>k_v$\+ \ \ \ //change epoch\\
$k_v = K$\\
set $B^v$ to its initial value \-\\
{\bf end If}\-\\
{\bf end If}\-\\
{\bf end While}\\
broadcast ($\{k_v,B^v\}$)\-\\
{\bf until}  \\
\end{tabbing}
\end{algorithm}

\section{Experiments}
\label{sec-Exp}
In this section we present an original application scenario which allowed us to implement and evaluate the performance of {\mgmc} under real mobile conditions and strong communication constraints. 
We designed a cycling race wireless sensor network, for assessing interactions between the riders and monitor the groups that can be formed during the race. A group is considered to split when the distance between the cyclists becomes greater than 20 meters.

We need to be aware of the topology changes at the timescale of the group motion, i.e. detect groups splitting or merging within a few seconds.
In addition, we have to take into account several mobile sinks which may appear or disappear in an uncontrolled manner, and we expect a refresh rate inferior to 1 Hz.

\begin{figure}[t!]
    \begin{center}
        \includegraphics[width=0.8\columnwidth]{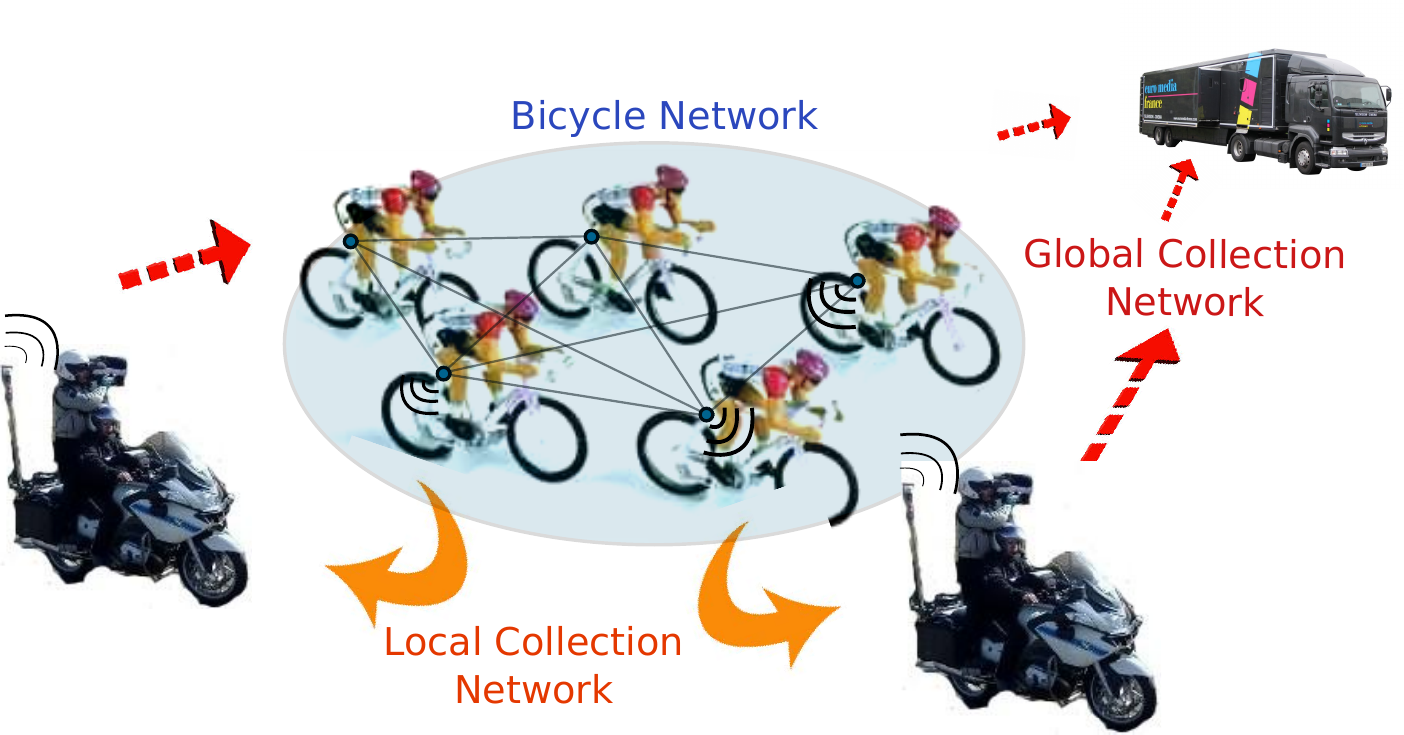}
    \end{center}
    \caption{Global infrastructure of the developed platform, illustrating the 3 data collection levels of the network.}
    \label{fig:infra}
\end{figure}

\newcommand{\fox}{\sc fox}
\newcommand{\lion}{\sc lion}
\newcommand{\ed}{\sc ed}
\newcommand{\rssi}{\sc rssi}

\subsection{Experimental setup}
The whole network infrastructure illustrated in Fig.\ref{fig:infra} is composed of three levels. The \emph{Bicycle Network} refers to the wireless sensor network formed by the nodes located on the bicycles; the \emph{Local Collection Network}  located on motorbikes surrounding the race aims at collecting the data shared by the \emph{Bicycle Network} and acts as a gateway, transmitting the collected information through a dedicated \emph{Global Collection Network} to a central sink located on a truck, where the data exploitation is performed. In this section we describe in details both the \emph{Bicycle Network} and the \emph{Local Collection Network}, but not the \emph{Global Collection Network}, which can be considered as a long-range RF tunnel, since the central point is located far away from the event.

\subsubsection{Bicycle Network}
Each bicycle is equipped with a HiKoB {\fox} sensor \cite{hikobfox}, fixed under the saddle as imposed for the cyclists' comfort. These sensors embed an Atmel AT86RF231 radio chipset embedding a IEEE 802.15.4 compliant PHY layer in the 2.4GHz ISM band, with CRC-16 error detection~\cite{atmel}; the integrated processor, used for the implementation of application algorithms and communication protocols, is a 32bits ARM Cortex M3 processor. As required for mobility, the FOX sensors run on batteries and embed a micro-SD storage facility, offering several hours of autonomy and storage capacity.
%

\subsubsection{Local Collection Network}
The system located on each motorbike is a HiKoB {\lion} router \cite{hikoblion}, which embeds a processor from the same family as the FOX sensor, and the same AT86RF231 radio chipset. It is connected to an external, high gain antenna, and directly powered on the motorbike. The received data is then formatted and transmitted through a USART on a standard asynchronous RS232 serial link with a 9600 bps bitrate before entering the Global Collection Network RF tunnel. 

\subsection{Calibration}
\label{sec:calib}
Before the real race, preliminary experiments have been performed to build a coarse distance estimator and to adapt and validate the communication protocols.

\subsubsection{Empirical distance evaluation}
\label{sec:calib_dist}
To fit with the requirements of the {\gmc} algorithm, every node must be able to estimate the distance with its 1-hop neighbours.

\begin{figure}[t!]
\centering
\includegraphics[scale=0.8]{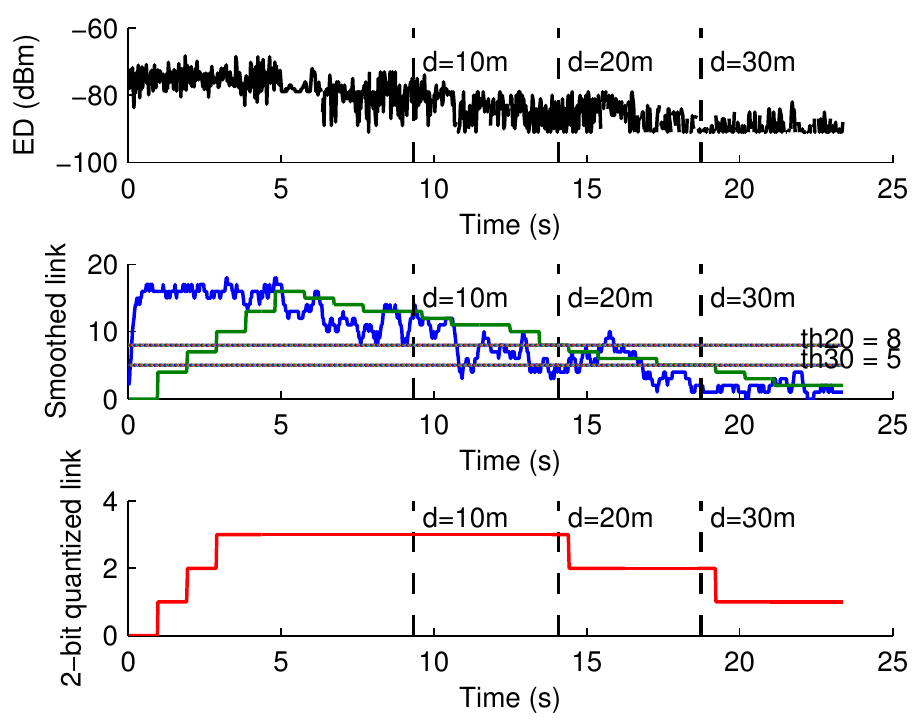}
\caption{Example of the smoothing and quantification process for link $l_{AB}$ using the retained parameters:
$W_1=10$, $W_2=5$, $\Delta_{t1}=1s$, $\Delta_{t2}=1s$, $th_{20}=8$, $th_{30}=5$. The black curve is the raw link measurement, blue and green are respectively the short and long term averages, while the red one is the result of quantification.}
\label{fig:smooth_dist}
\end{figure}

We describe in \cite{mobihoc} the platform we developed to periodically assess the channel between {\fox} nodes under controlled cycling conditions, using the Energy Detection measurement given by the radio; we also provide explanation on the empirical {\ed} smoothing and quantization method we propose to roughly estimate the distance between the nodes.
An illustration of this process is given in Fig.~\ref{fig:smooth_dist}.

\subsubsection{Algorithm and Protocol Calibration}
\label{sec:prot_calib}
The practical algorithm we implemented is the fully extended {\mgmc} Algorithm (algorithm \ref{algo3}), with an external beacon periodically sent by the {\lion} routers for new epoch propagation.

If we now focus on the communication design, we need to ensure the reception of at least 10 packets per node per second for neighbours in a close communication range, to obtain a correct distance estimation.
Our algorithm relies on a random organisation protocol, which is justified by the simple fact that in those conditions synchronised protocols may be very difficult to implement, and not easy to adapt to topology changes. We experimentally fixed the parameter $T=70ms$ as this value ensures a globally fair reception rate around 15 packets per second per node for a static experiment, i.e. all nodes on a table, giving a maximum  convergence time of 200 ms. To limit collisions, the implemented communication protocol is based on {\sc csma/ca} without acknowledgement, $T_{max}$ being considered as a random backoff, each node freezing the decrease of $T_{max}$ when sensing the channel busy.

The period duration was set to 400 ms to ensure the convergence in moving conditions, i.e. taking into account faulty links.
All the received packets were locally stored on micro-SD cards, as well as additional information, such as the {\ed} value measured for each packet, the local reception timestamp, the number of packets sent every second, and the amount of erroneous (wrong-CRC) packets received.

\subsection{Experimental Conditions and Results}
\label{sec-results}
We first explain the experimental setup before presenting some interesting results extracted from the collected data. The performance of our algorithms are presented for a stable group but also in a situation where merge-and-split variations occur, as requested for the application.

\subsubsection{Experimental Conditions}
\begin{figure}[t]
\centering
\includegraphics[scale=1]{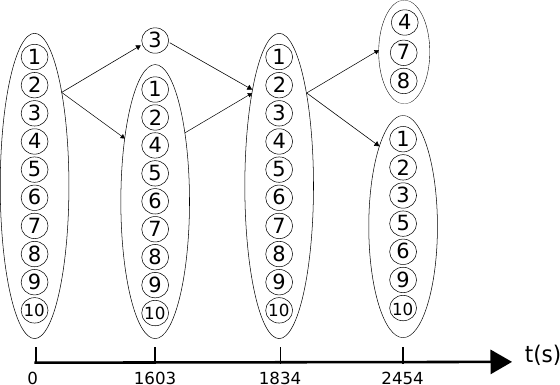}
\caption{Global progression of the bicycle network during the experiment. We represent here the three main events that were detected and the instants of detection, i.e. isolated racer moving away and joining again the pack, and the pack splitting in two parts.}
\label{fig:prog_expe}
\end{figure}

The experiment was conducted with a group of ten racers, using the global infrastructure represented in Fig.\ref{fig:infra}, in the region of Paris, for about $1$ hour, which allowed us to test the reliability of our algorithms and the whole communication platform over time. The circuit was a $2 km$ loop, in a semi-urban environment, i.e. with both buildings and rather clear areas. As we explained in \ref{sec:prot_calib}, for the whole duration of the race every node stored all the packets it received, plus additional data, on its micro-SD card. These data provided an important database from which the performance of our algorithms was studied. We focus here only on two major racing situations, the first corresponding to a stable group and the second to a dynamic splitting. Indeed, the race started with a long period during which all the racers were forming a unique pack. During this period, one racer shortly moved away before joining in again. After that move, the group split in two sub-groups until the end of the event, one motorbike following each formation. The progress of the race was extracted from the stored data, and is described in Fig.\ref{fig:prog_expe}, which validates our platform from the application point of view.

\subsubsection{Stability and Performance}
\begin{figure}[t]
\centering
\includegraphics[scale=0.8]{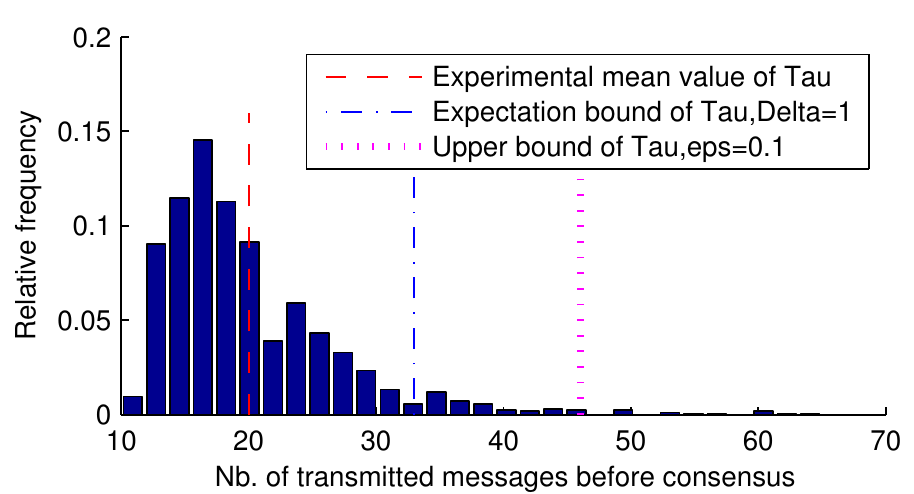}
\caption{Experimental distribution obtained for the number of messages needed before convergence $\tau$, $N=10$, $\Delta_C\leq 2$, compared to the theoretical bounds given in equations~(\ref{theorem1}) and~(\ref{theorem2})with $\Delta=1$ and $\epsilon=0.1$.}.
\label{fig:expe_distrib}
\end{figure}

We will focus here on the first part of the experiment, for $t\in [0;1603s]$, during which all the cyclists are riding together, without controlling more their motion, i.e. relative positions and distances may vary. This first step is important to estimate the performances on the full graph before focusing on group splitting.
First of all, we obtained an average packet loss over all this period of 22\%, which is non negligible but seems reasonable given the important traffic and the transmission conditions (motion, bikers acting as obstacles, channel instability\ldots).
It is now interesting to focus on the number of messages exchanged before reaching consensus, and compare it to the theoretical bounds. According to our measurements, during this period the graph diameter (taking into account the distance criterion) is low, $1\leq \Delta_C\leq 2$, which means that all nodes are almost direct mates.
We study the behavior of $\tau$ introduced in section 3.1, which is the number of messages exchanged before algorithm convergence.
Fig.~\ref{fig:expe_distrib} shows the  experimental distribution of $\tau$, 
which validates the theoretical bounds given in equations~(\ref{theorem1}) and~(\ref{theorem2}) in experimental conditions.


\subsubsection{Dynamic Splitting}
\begin{figure}[t]
\centering
\includegraphics[scale=0.5]{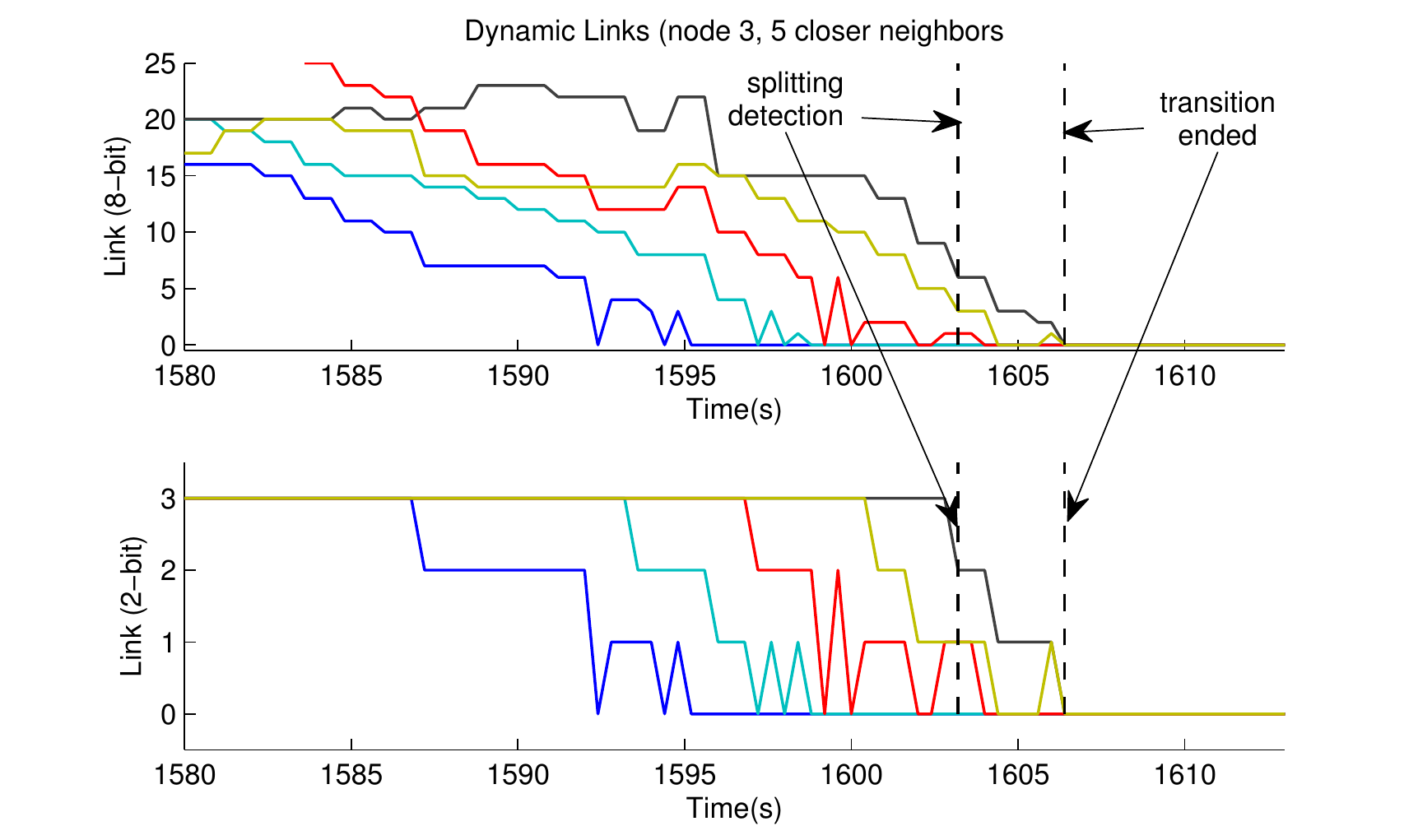}
\caption{Dynamic link behaviour at node 3 when getting away from the pack (5 closer links). Splitting is detected when all nodes have a weaker proximity index with node 3. During the transition phase, both groups share a weaker link, after what they are out of each other's communication range.}
\label{fig:split}
\end{figure}

The convergence of our algorithm being validated for a static graph, we now observe with more details how it performs on a dynamic topology by exploiting the distance estimation in real conditions. From the application point of view, an interesting dynamic scenario is typically when the pack comes to split in case a breakaway occurs.
We focus on the first event described in Fig.\ref{fig:prog_expe} when node $3$ moves away rapidly from the rest of the racers, and we analyze the distance indicators that were computed according to the method described in \ref{sec:calib_dist}. Fig.\ref{fig:split} shows the evolution of the links between node $3$ and its $5$ closer neighbours, during that splitting phase.
The first observation is that in practice our distance index
remains stable and monotonous despite the fact that,  having the pack forming a single line, every racer behaves as a communication obstacle,
which adds uncertainty on the link measurement for nodes
located at several hops.
The second observation is that we can observe a short transition phase, during which two groups are detected but with a weak link between them, before being considered fully disjoint. This expected behavior tends to validate the implemented methods in a real life uncontrolled scenario.

\section{Conclusion}
We described in this paper efficient algorithms for group consensus, self-organised and adapted to mobile applications, capable to detect fast topology changes. We provided theoretical bounds on their convergence performances, which were experimentally validated through a challenging application scenario, for which all the functionalities were implemented.
One limitation is that in the context of a cycling race, a high proportion of messages are transmitted for the computation of the distance estimation, due to the lack of accuracy of the RSSI measurement.
The lack of accurate technological solutions for distance estimation may not be an issue in the future, with the apparition of UWB chipsets implementing time-based distance measurements, which are less dependent to the communication environment.
This important network load must be taken into account according to the number of nodes communicating together, to avoid channel saturation and unefficient communications due to a high number of collisions.
In the case of a growing number of nodes, it would be necessary to implement mechanisms that control the nodes' communication range, e.g. by adapting the transmission power to the density.
As the packet size is also proportional to the number of nodes, the use of adaptive data compression methods could be of interest, in order to reduce the data exchanged without degrading the quality of the measurements.

\section*{Acknowledgement}
We wish to thank the company Euromedia France for funding this project, and for their technical collaboration for the realization of the experiment, giving us the opportunity to design an original experimental scenario with the complete data collection infrastructure. We are also grateful to the company HiKoB, for their important technical support with the design of the platform.

\bibliographystyle{abbrv}
\bibliography{Bodynets}

\end{document}

%% file: staticConsensus.tex
In this section, we present the {\em mobile group consensus algorithms} designed for the cyclist race. Their implementation in a real experimentation are provided in (section~\ref{sec-Exp}). We first present the basic algorithm 
dealing with a binary notion of neighborhood and with static nodes, which means that the connectivity between nodes
does not change over time. We then introduce the {\em mate} concept of ``close'' neigbhour (section~\ref{sec-extension}) and finally present a solution to manage dynamic networks (section~\ref{sec:period}).

\subsection{Static Group Consensus Algorithm}
\label{sec-static}

Our first  algorithm is a straight extension of a Max-Consensus Algorithm presented in~\cite{Iutzeler2012}. Iutzeler et al. present a Max-Consensus Algorithm for non-mobile nodes in which  each node should become aware of the maximal value held by all the nodes. We  extend this algorithm so that each node should become aware of {\em all} the values held by all other nodes. 

The static wireless sensor network is modeled as an undirected graph
$G=(V,E)$.  $V$ is the set of sensors ($|V|=N$)  and $E$ the set of sensor connections.  
We denote $\Delta_G$ the diameter of graph $G$. 
In this section \newcommand{\iid}{i.i.d}, each link is supposed to be error-free and
constant in time without collision.

\newcommand{\sgc}{\sc sgc}

The Static Group Consensus ({\sgc}) Algorithm aims at allowing each
node in a graph to obtain the list of its 
connected components (i.e. the set of nodes which are connected to
him through a multi-hop path).
Each node $v$ possesses an internal Boolean $N$-vector $B^v$ containing
the information of the nodes it can reach. After running the {\sgc}
algorithm, $B^v_i$ is equal to $1$ if nodes $i$ and $v$ are in the same
connected component.

\newtheorem{algorithm}[equation]{Algorithm}

The {\sgc} algorithm is presented in Algorithm~\ref{algo1}, $Max$ is
the component-wise maximum operation 
and we give below an estimation of the convergence time.

\begin{algorithm}
\label{algo1}
{\sc Static Group Consensus Algorithm }
\begin{tabbing}
==\= ===\= ===\= ===\= ===\kill
(Executed on each node $v$)\+ \\
Initialize $B^v$ with a $1$ at component $v$ and 0 elsewhere\\ 
{\bf repeat}\+\\
$T_{max}=random(0-T)$\\
{\bf While}  not expired backoff $T_{max}$ \+ \\
receive($B^j$) from node $j$\\
$B^v$=$Max(B^v,B^j)$\-\\
{\bf end While}\\
broadcast ($B^v$)\-\\
{\bf until}  finished
\end{tabbing}
\end{algorithm}

\newcommand{\setone}{\bold{1}}

The {\sgc} algorithm convergence rate has been studied in~\cite{mobihoc}. It is defined as the number of rounds $\tau$ needed to reach the state where all $B^v$ are equal to $\setone$, and is such that

\addtocounter{equation}{-1}
\begin{equation}
\label{theorem1}
\mathbb{E}[\tau] < N\Delta_G(1+log(N))
\end{equation}

It is also shown in~\cite{mobihoc} that an upper bound on $\tau$ can be given with an arbitrary probability: $1-\epsilon$
\begin{equation}
\label{theorem2}
\tau < N\Delta_G\left(log(N)+log\left(\frac{\Delta_G}{\epsilon}\right)\right)
\end{equation}

It is true noting that these theoretical bounds assume a pure random selection at each turn with not any priority, while the implementation exploits a random backoff mechanism which may increase the fairness. 

%
%
%

\newcommand{\rbm}{\sc RBM}
\newcommand{\rgg}{\sc RGG}




%% file: Bodynets_short.bbl
\begin{thebibliography}{10}

\bibitem{atmel}
{AT}86{RF}231.
\newblock \url{www.atmel.com/devices/at86rf231.aspx}, 2012.

\bibitem{hikobfox}
{H}i{K}o{B} {FOX} sensor.
\newblock \url{www.hikob.com/hikob-fox}, 2012.

\bibitem{hikoblion}
{H}i{K}o{B} {LION} sensor.
\newblock \url{www.hikob.com/hikob-azure-lion}, 2012.

\bibitem{Gossip2}
T.~Aysal, M.~Yildiz, A.~Sarwate, and A.~Scaglione.
\newblock Broadcast gossip algorithms for consensus.
\newblock {\em IEEE Transactions on Signal Processing}, 57(7):2748--2761, 2009.

\bibitem{Gossip}
S.~Boyd, A.~Ghosh, B.~Prabhakar, and D.~Shah.
\newblock Randomized gossip algorithms.
\newblock {\em IEEE Transactions on Information Theory}, 52(6):2508--2530,
  2006.

\bibitem{Cortes2008}
J.~Cortes.
\newblock Distributed algorithms for reaching consensus on general functions.
\newblock {\em Automatica}, 44(3):726 -- 737, 2008.

\bibitem{Iutzeler2012}
F.~Iutzeler, P.~Ciblat, and J.~Jakubowicz.
\newblock Analysis of max-consensus algorithms in wireless channels.
\newblock {\em IEEE Transactions on Signal Processing}, 60(11):6103--6107,
  2012.

\bibitem{kar2009distributed}
S.~Kar and J.~M. Moura.
\newblock Distributed consensus algorithms in sensor networks with imperfect
  communication: Link failures and channel noise.
\newblock {\em IEEE Transactions on Signal Processing}, 57(1):355--369, 2009.

\bibitem{kermarrec2003probabilistic}
A.-M. Kermarrec, L.~Massouli{\'e}, and A.~J. Ganesh.
\newblock Probabilistic reliable dissemination in large-scale systems.
\newblock {\em IEEE Transactions on Parallel and Distributed Systems},
  14(3):248--258, 2003.

\bibitem{mobihoc}
M.~Lauzier, A.~Fraboulet, J.~Gorce, and T.~Risset.
\newblock Distributed group consensus algorithms for mobile wireless sensor
  networks.
\newblock Technical Report RR-8518, INRIA, 2014.

\bibitem{marechal2010joint}
N.~Mar{\'e}chal, J.-M. Gorce, and J.~Pierrot.
\newblock Joint estimation and gossip averaging for sensor network
  applications.
\newblock {\em IEEE Transactions on Automatic Control}, 55(5):1208--1213, 2010.

\bibitem{Boyd}
L.~Xiao, S.~Boyd, and S.~Lall.
\newblock A scheme for robust distributed sensor fusion based on average
  consensus.
\newblock In {\em Proceedings of the 4th International Symposium on Information
  Processing in Sensor Networks}, IPSN '05, Piscataway, NJ, USA, 2005. IEEE
  Press.

\end{thebibliography}
